\documentclass[aps,prd,superscriptaddress,long,notitlepage,balancelastpage,nofootinbib,floatfix,twocolumn]{revtex4-1}
\pdfoutput=1

\usepackage{amsmath,mathtools,amssymb,amsbsy,amsthm,amsxtra,overpic,bbm,epsfig,subfigure,url,graphics,graphicx,multirow,booktabs,tabularx}
\usepackage{mathrsfs}
\usepackage{graphicx,soul}
\usepackage{comment}
\usepackage{float}
\usepackage{enumitem}
\usepackage{slashed}
\usepackage{tikz}
\usepackage{bm}
\usepackage{slashed,stmaryrd}
\usepackage{bold-extra}
\usepackage{verbatim}
\usepackage{fixmath}

\setlist{nolistsep}

\definecolor{MyDarkBlue}{rgb}{0.1, 0.1, 0.8} 
\definecolor{SBlue}{rgb}{0.2, 0.4, 0.7} 
\definecolor{MyLightBlue}{rgb}{0.22,0.51,0.9}
\definecolor{MyGreen}{rgb}{0.0, 0.5, 0.0}
\definecolor{BrickRed}{rgb}{0.8, 0.25, 0.33}
\usepackage{hyperref}
\hypersetup{colorlinks, citecolor=SBlue,linkcolor=MyDarkBlue, urlcolor=BrickRed}

\newcommand{\Xe}{$^{136}$Xe}

\begin{document}
	\title{Inference of neutrino nature and Majorana CP phases from ${\bf 0}\mathbold{\nu\beta\beta}$ decays \\ with inverted  mass ordering}
	\author{Guo-yuan Huang}
	\email{guoyuan.huang@mpi-hd.mpg.de}
	\affiliation{Max-Planck-Institut f{\"u}r Kernphysik, Saupfercheckweg 1, 69117 Heidelberg, Germany} 
	
	\author{Newton Nath} 
	\email{newton.nath@ba.infn.it}
	\affiliation{Istituto Nazionale di Fisica Nucleare,   Via  Orabona  4,  70126  Bari, Italy}

	\begin{abstract}
		\noindent
		Whether the neutrino mass ordering is normal or inverted remains an experimentally open issue in neutrino physics. 
	The knowledge of neutrino mass ordering has great importance for neutrinoless double-beta ($ 0\nu\beta\beta$) decay experiments, which can establish the nature of massive neutrinos, i.e., whether they are Dirac  or Majorana fermions.
		Recently, the KamLAND-Zen 800 measurement has reached for the first time the parameter space of the inverted ordering with a vanishing lightest neutrino mass.
		By assuming the inverted ordering, we attempt to derive the physical information of the neutrino nature and Majorana CP phases from a negative or positive observation of $ 0\nu\beta\beta$ decays in the near future. Moreover, the possibility of extracting the nuclear matrix element in the case of a positive observation is also examined.
		To avoid the ambiguity from unknown priors of neutrino masses, we adopt the maximum likelihood method instead of the Bayesian approach usually considered in previous works.
		%
	\end{abstract}
	
	\maketitle
	
	\section{Introduction}
	\noindent
	The nature of massive neutrinos, whether they are Majorana or Dirac fermions, has been a longstanding quest in neutrino physics~\cite{Majorana:1937vz}. 
	At present, the most feasible process that can uncover the neutrino nature is the neutrinoless double-beta  ($ 0\nu\beta \beta $) decay (for recent reviews, see Refs.~\cite{Agostini:2022zub,Cirigliano:2022oqy,Dolinski:2019nrj,DellOro:2016tmg,Vergados:2016hso}). 
	Such a process takes place through violating the lepton number by two units~\cite{Furry:1939qr},
	i.e., $(Z, A) \to (Z+2, A) + 2 e^{-}$, where   $Z$ and $A$ represent the atomic and  mass numbers of a nucleus, respectively.
	For a given $0\nu\beta\beta$ decay process, the half-life of the isotope of concern can be calculated by~\cite{Rodejohann:2011mu}
	\begin{equation}\label{eq:HalfLife}
	\dfrac{1}{T^{0\nu}_{1/2}} = G^{}_{0\nu} \cdot \left|{ M}^{}_{0\nu}\right|^{2} \cdot \frac{\left|m_{e e} \right|^{2} }{M^2_e }\; ,
	\end{equation}
	where  $G_{0\nu}$ denotes the kinematic phase-space factor, $ M_{0\nu} $ is the nuclear matrix element (NME), and $M^{}_{e} \approx 0.51~{\rm MeV}$ is the electron mass. 
	In the standard three-flavor paradigm, the half-life of $0\nu\beta\beta$ decays is controlled by the  effective Majorana neutrino mass,
	%
	\begin{eqnarray}\label{eq:mbb}
	{m}^{}_{ee} &  \equiv & m^{}_1 \cos^2 \theta^{}_{13} \cos^2 \theta^{}_{12} e^{{\rm i}\rho} + m^{}_2 \cos^2 \theta^{}_{13} \sin^2 \theta^{}_{12} \notag  \\ & &+ \ m^{}_3 \sin^2 \theta^{}_{13} e^{{\rm i}\sigma} \; , 
	\end{eqnarray}
	where $m^{}_i$ (for $i = 1, 2, 3$) stand for the absolute masses of three neutrinos, $\{\theta^{}_{12},\theta^{}_{13} \}$ are the leptonic mixing angles with the standard parametrization, and $\{\rho, \sigma\}$ are the Majorana CP phases.
	As the overall CP phase of $m^{}_{ee}$ is non-physical, it is a convenient choice to assign two Majorana CP phases to the lightest and the heaviest neutrino states~\cite{Xing:2014yka,Xing:2015uqa,Xing:2015kfr}, such that only one phase parameter will survive when the lightest neutrino mass is vanishing.

	The importance of $0\nu\beta\beta$ decay searches for neutrino physics is well recognized~\cite{Dorame:2011eb,BhupalDev:2013ntw,Xing:2015zha,Zhang:2015kaa,Benato:2015via,Lisi:2015yma,Xing:2016ymd,Ge:2016tfx,Agostini:2017jim,Nath:2018zoi,Penedo:2018kpc,Cao:2019hli,Ge:2019ldu,Huang:2020mkz,Huang:2020kgt,Biller:2021bqx,Hu:2021ziw,Agostini:2021kba,Lisi:2022nka}, as it can answer: (i) the nature of neutrinos, (ii) the Majorana CP phases, (iii) the absolute scale of neutrino masses, and (iv) the neutrino mass ordering.
	While the neutrino masses can be better probed by other observations (e.g., cosmology and oscillation experiments), the $0\nu\beta\beta$ decay search seems to be the only reliable way towards
	the neutrino nature and the Majorana CP phases.

	\begin{figure*}[t!]
		\begin{center}
			\includegraphics[width=0.4\textwidth]{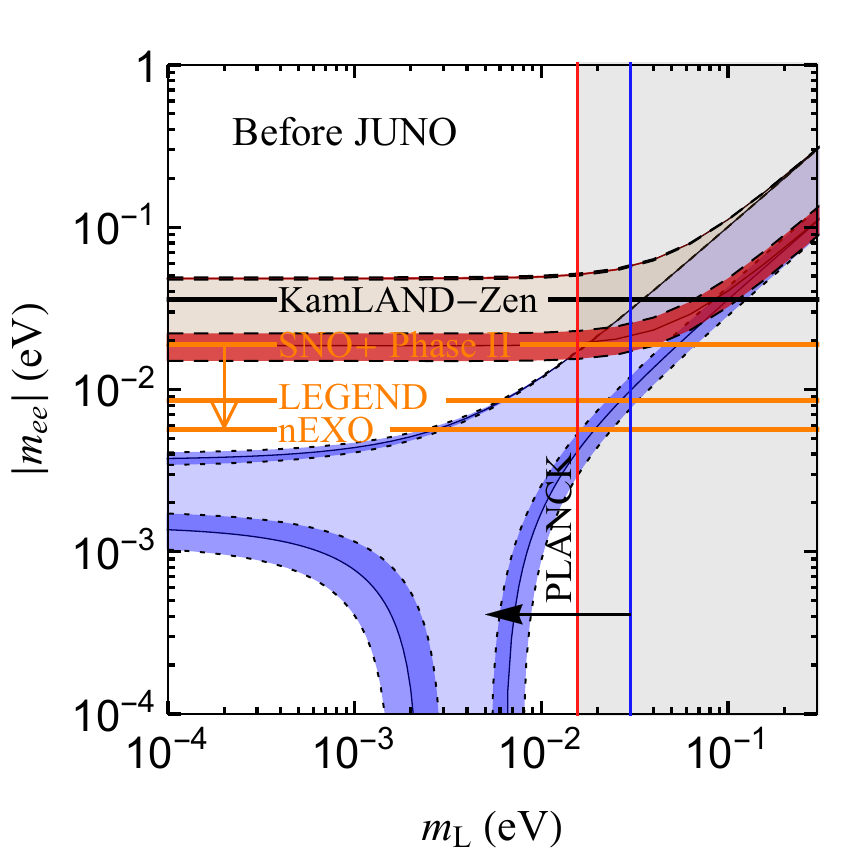}
			\includegraphics[width=0.4\textwidth]{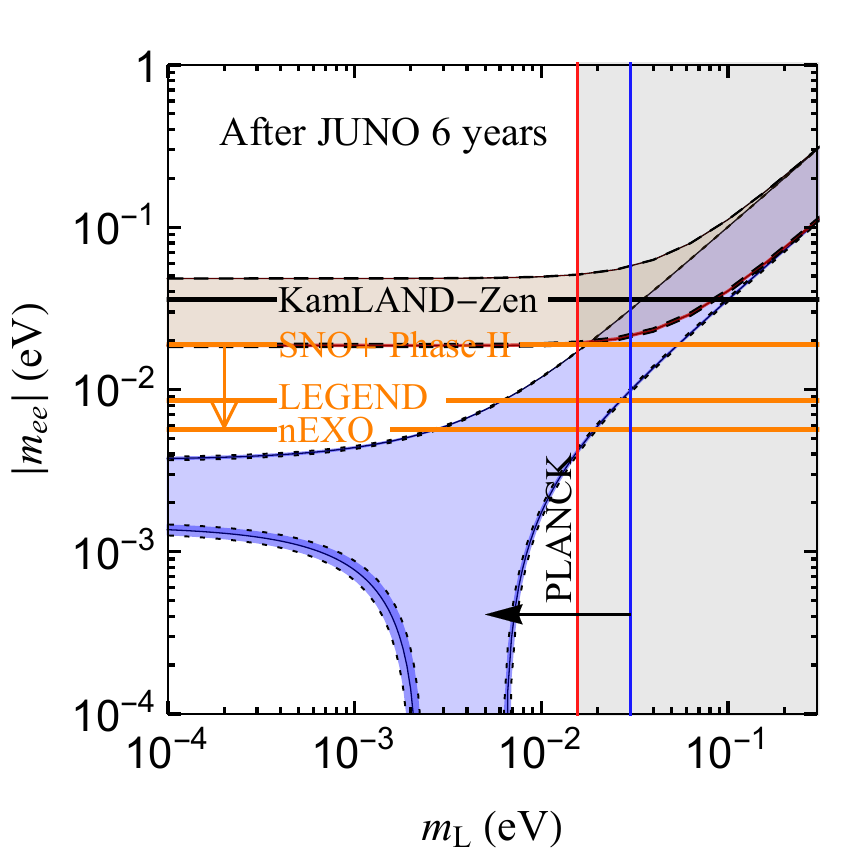}
			\includegraphics[width=0.4\textwidth]{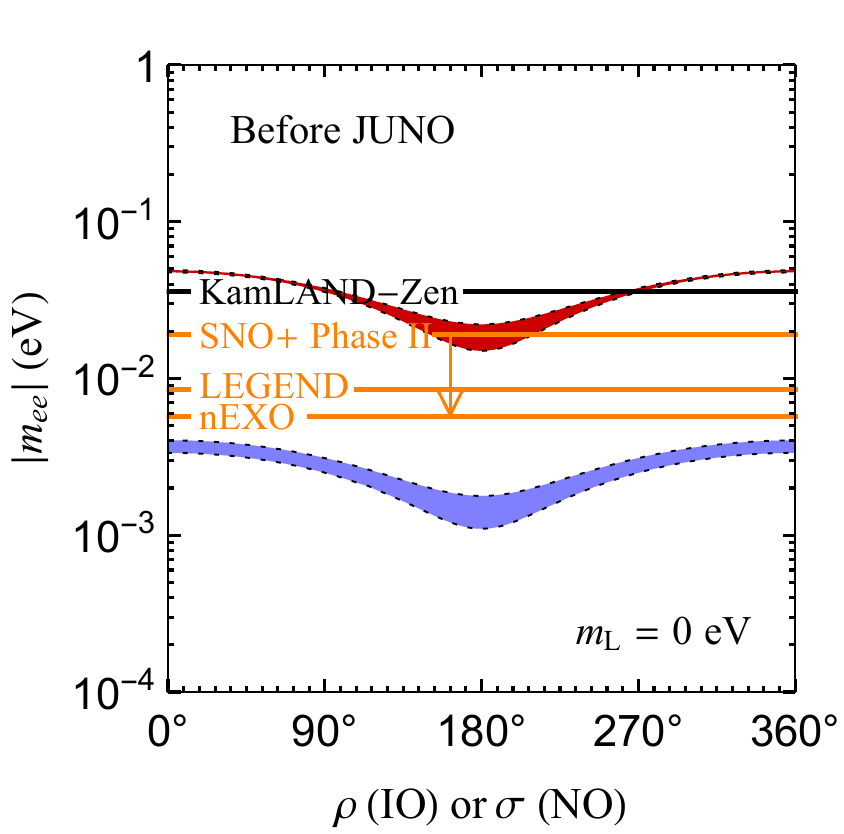}
			\includegraphics[width=0.4\textwidth]{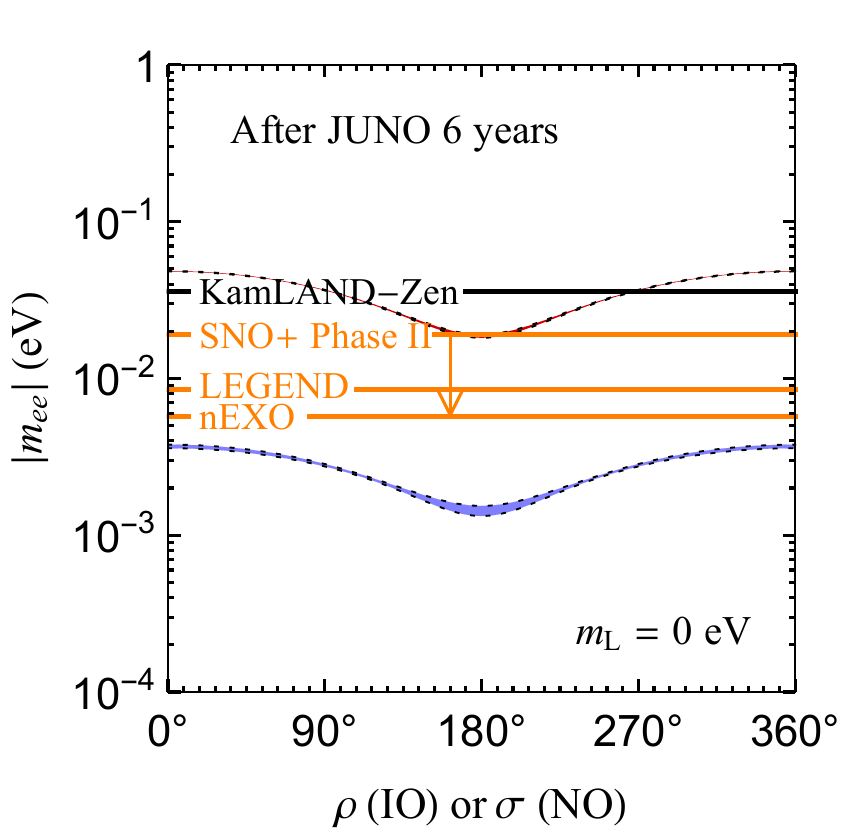}
		\end{center}
		\vspace{-0.3cm}
		\caption{The effective neutrino mass $|m^{}_{ee}|$ as a function of the lightest neutrino mass (upper two panels) or the Majorana CP phase (lower two panels). 
			The darker blue (for NO) and red (for IO) regions signify the $3\sigma$ uncertainties ($\chi^2_{\rm osc} = 9$) caused by neutrino oscillation parameters, i.e., $\theta^{}_{13}$, $\theta^{}_{12}$, $\Delta m^{2}_{\rm sol}$ and $\Delta m^2_{\rm atm}$, and the lighter blue and red regions stem from the unknown Majorana CP phases $\rho$ and $\sigma$.
			The left two panels are calculated by adopting the NuFIT5.1 global-fit results, while the right two panels give the prospects of $3\sigma$ uncertainties after JUNO running for 6 years. 
			For comparison, the KamLAND-Zen limit on $|m^{}_{ee}|$ and projections  of several future experiments (SNO+ Phase II~\cite{Andringa:2015tza}, LEGEND~\cite{LEGEND:2021bnm} and nEXO~\cite{Albert:2017hjq}) at 90\% confidence level (C. L.) are shown as the horizontal black and orange lines (with the {\it most optimistic} NME).
			For the lower two panels,  the lightest neutrino mass $m^{}_{\rm L}$ is taken to be vanishing. }
		\label{fig:vissani}
	\end{figure*}

	In order to establish the relation between the observable $T^{0\nu}_{1/2}$ and the quantity of our interest ${m}^{}_{ee}$, it is necessary to have a robust theoretical calculation of NMEs  (see, e.g., Refs.~\cite{Engel:2016xgb,Davoudi:2020ngi,Yao:2021wst} for recent reviews).
	Interestingly, it was pointed out recently that a short-range effect at the leading order can potentially increase the NME by $40\%$~\cite{Cirigliano:2018hja,Cirigliano:2020dmx,Wirth:2021pij,Jokiniemi:2021qqv}.
	Recent developments, especially those in the \textit{ab initio} approach~\cite{Pastore:2017ofx} and the lattice-QCD calculation~\cite{Beane:2010em}, might allow the nuclear physicists to have a more robust evaluation of nuclear matrix elements with accessible errors. 
	With that being achieved, one is able to derive much more information for the neutrino nature and CP phases from the $0\nu\beta\beta$ decay data.
	In the pessimistic  case, the anarchic status of NMEs based on various phenomenological evaluations will persist, if these advances encounter possible computational bottleneck when adapting to the heavy nuclei.
	One may, in such a case, go the other way around to constrain NMEs with $0\nu\beta\beta$ decay data, in the assumption that our understanding of the neutrino mass pattern is further improved in the future.

	Depending on whether neutrino masses are normal ordering (NO) or inverted ordering (IO), the predictions of $|m^{}_{ee}|$ differ significantly.
	In fact, it has been recognized that this feature allows us to distinguish NO and IO based on the $0\nu\beta \beta$ measurement alone~\cite{Zhang:2015kaa}.
	However, the determination of mass ordering, with a high significance, is already achievable by upcoming neutrino oscillation experiments such as JUNO~\cite{JUNO:2015zny}, DUNE~\cite{DUNE:2020jqi} and KM3NeT (ORCA) \cite{KM3NeT:2021ozk} in the foreseeable future.
	If the mass ordering turns out to be normal, we do not expect any $0\nu\beta\beta$ decay signals  with a high probability in the near future.
	A positive signal would mean either new physics contributions \cite{Goswami:2005ng,Goswami:2007kv,Barry:2011wb,Li:2011ss,BhupalDev:2013ntw,Girardi:2013zra,Guzowski:2015saa,Giunti:2015kza,Ge:2017erv,Liu:2017ago,Cepedello:2018zvr,Huang:2019qvq,Vishnudath:2019eiu,Graf:2020cbf,Majumdar:2020xws,Ghosh:2020fes,Fang:2021jfv,Huang:2021kam,Gariazzo:2022ahe} in $|m^{}_{ee}|$ or an unusual cosmology with large neutrino masses~\cite{Davoudiasl:2018hjw,Huang:2021zzz,Esteban:2021ozz,Alvey:2021xmq}.
	If the mass ordering turns out to be inverted, in contrast to the current global-fit preference~\cite{Capozzi:2021fjo, deSalas:2020pgw, Esteban:2020cvm}, we will instead have a considerable discovery probability.
	A positive signal would not only mean neutrinos are Majorana fermions but also allow for constraining the Majorana CP phases and NMEs.
	A negative result would then favor Dirac neutrinos, contrary to the likes of most neutrino model builders.

	In the upper two panels of Fig.~\ref{fig:vissani}, we give $|m^{}_{ee}|$ as a function of the lightest neutrino mass $m^{}_{\rm L}$ for NO (in blue) and IO (in red). 
	The optimistic limit of KamLAND-Zen, $ |m_{ee} | < 36~{\rm meV}$  at 90\% C.~L., has touched for the first time the IO prediction with a vanishing $m^{}_{\rm L}$~\cite{KamLAND-Zen:2022tow}. The next-generation proposals with a  sensitivity of $|m^{}_{ee}| \sim \mathcal{O}(10~{\rm meV})$ at 90\% C.~L., such as CUPID $(8.4–14$ meV)~\cite{CUPID:2022wpt}, JUNO-$\beta\beta$ $(5 – 12$ meV)~\cite{Zhao:2016brs}, KamLAND2-Zen ($< 20$ meV)~\cite{Nakamura:2020szx}, LEGEND 1000 ($8.5-19.4$ meV)~\cite{LEGEND:2021bnm}, nEXO ($5.7-17.7$ meV)~\cite{Albert:2017hjq}, NEXT 1T~\cite{NEXT:2020amj}, PandaX-III 1k ($20-46$ meV)~\cite{Chen:2016qcd} and ${\rm SNO}+$ Phase II ($19-46$ meV)~\cite{Andringa:2015tza}, are expected to cover the full parameter space of IO (with the most optimistic NME), while a large fraction of NO parameter space remains untouched.
	On the other side,
	the cosmological observations of Planck~\cite{Aghanim:2018eyx} already pushed $m^{}_{\rm L}$ to the region where NO and IO predictions start to separate. 
	In the lower panels of Fig.~\ref{fig:vissani}, we present $|m^{}_{ee}|$ with respect to the Majorana CP phases. One can see from Eq.~\eqref{eq:mbb} that $|m^{}_{ee}|$  depends on only one CP phase when the lightest neutrino mass is vanishing, i.e., $m_1 \rightarrow 0 $ ($m_3 \rightarrow 0 $) for NO (IO). 
	This specific example  shows visually how a measurement of $|m^{}_{ee}|$ is able to reflect the information of CP phases. Note that in the later numerical analysis, we do not restrict the lightest	neutrino mass to be vanishing.

	The errors on neutrino oscillation parameters that enter as inputs in $|m^{}_{ee}|$, are still large even though their measurements have entered the precision era.
	In this regard, JUNO can not only pin down the crucial neutrino mass ordering but also provide a precise measurement of $\{\theta^{}_{12}, \Delta m^{2}_{\rm sol}, \Delta m^2_{\rm atm}\}$ after 6 years of running (see Table~\ref{Tab:OscParamAccuracy}).
	The induced $3\sigma$ uncertainty in $|m^{}_{ee}|$ is shown as the darker blue and red regions in the left (or right) panel of Fig.~\ref{fig:vissani}, using the current global-fit (or JUNO 6-year) results of neutrino oscillation parameters.
	One can see from the right panel a significant improvement in the predicted regions of  $|m^{}_{ee}|$. 
	The rest of the uncertainty in $|m^{}_{ee}|$ (in lighter blue and red) is attributed to the completely unknown Majorana CP phases.
	Similar observations have also been made in Refs.~\cite{Ge:2015bfa,Cao:2019hli,Ge:2019ldu}.
	
	In the assumption of NO,  a meV sensitivity to $|m^{}_{ee}|$ is required to extract possible information of Majorana CP phases~\cite{Xing:2015zha,Xing:2016ymd,Ge:2016tfx,Penedo:2018kpc,Ge:2019ldu,Cao:2019hli,Huang:2020mkz}, which is, however, far beyond the capability of the next-generation experiments in the near future.
	One would wish that IO is true from a practical point of view, as the full parameter space of IO will be covered with a $10~{\rm meV}$ sensitivity in the not-too-distant future.
	Standing in this paradoxical situation, we ask ourselves what we can infer quantitatively if the next-generation  $0\nu\beta\beta$ decay experiments come out with positive or null signals, assuming JUNO data point towards IO along with improved measurements of oscillation parameters.

	We structure the article as follows. In Sec.~\ref{sec:StatFrame}, we describe the statistical framework for our purpose.
	Our quantitative results are presented in Sec.~\ref{sec:Result}, including the implications for the nature of neutrinos (Sec.~\ref{sec:DorM}), the Majorana CP phases (Sec.~\ref{sec:MajCP}) in the case of Majorana neutrinos as well as the NME (Sec.~\ref{sec:NME}).  
	Finally, we conclude in Sec.~\ref{sec:conclusion}.
	%

	\section{Statistical framework}\label{sec:StatFrame}
	\noindent
	While it is straightforward to  establish the one-to-one analytical correspondence between $T^{0\nu}_{1/2}$ and $|m^{}_{ee}|$ for any specified NMEs, a more quantitative analysis can only be made within a given statistical framework.
	In the present work, we use the frequentist approach to test different models and to estimate model parameters.

	In contrast, previous works were mainly built on the Bayesian approach~\cite{Zhang:2015kaa,Benato:2015via,Ge:2016tfx,Agostini:2017jim,Ge:2017erv,Huang:2019qvq,Ge:2019ldu,Huang:2020mkz}, which, however, severely relies on the unknown priors of model parameters.
	The prior of a parameter in Bayesian statistics should be assigned according to its physical origin.
	The most pronounced ambiguity for $0\nu\beta\beta$ decays comes from the prior of neutrino masses. 
	Indeed, the origin of neutrino masses is not yet known, 
	and different \textit{ad hoc} choices, usually depending on one's belief, can yield very distinct results~\cite{Zhang:2015kaa,Agostini:2017jim,Huang:2019tdh,Huang:2019qvq,Huang:2020mkz,Gariazzo:2022ahe}.
	Moreover, there is also ambiguity in the priors of angles and phases. For instance, assigning a flat prior to $\rho$ or $\sin{\rho}$ will produce different interpretations.
	The absence of a trustworthy prior choice leads to the considerations of objective (uninformative) priors when evaluating a specific parameter of concern~\cite{Heavens_2018,Gariazzo:2022ahe}. 
	However, it is not a trivial task to find such priors, and the priors objective for the estimation of one parameter might be biased for another. 
	To avoid the ambiguity of the prior choice, we hence adopt the maximum likelihood method in line with frequentist's taste.

	%
	
	\begin{table*}[t!]
		\centering
		\begin{tabular}{lllllll}
			\hline\hline
			Oscillation parameters & Capozzi et al.\text{    } & de Salas  et al.\text{    } & NuFIT 5.1\text{    } & JUNO\text{    } & DUNE\text{    } & HK\text{    } \\
			\hline
			$\bm{\Delta m^2_{\rm sol}}/(10^{-5}~\mathrm{eV}^2) $ & 7.36 (2.3 \%) & 7.50 (2.6 \%) & 7.42 (2.7 \%) &  0.3 \% & $\cdots$ & $\cdots$  \\
			%
			%
			$\bm{|\Delta m^2_{\rm atm}|}/(10^{-3}~\mathrm{eV}^2) $ & 2.46 (1.1 \%)  & 2.45 (1.1 \%) & 2.49 (1.1 \%)  & 0.2 \%  & 0.5 \% &  0.6\%\\
			$\bm{{\rm sin}^2 \theta_{12}}/10^{-1}$ & 3.03 (4.5 \%) & 3.12 (5.1 \%) & 3.04 (4.1 \%)  & 0.5 \%  &  $\cdots$ & $\cdots$  \\
			$\bm{{\rm sin}^2 \theta_{13}}/10^{-2}$ & 2.23 (3.1 \%) & 2.23  (3.0 \%) & 2.24 (2.8 \%)  &  12\% & 7 \%  &  $\cdots$ \\
			$\sin^2 \theta_{23}/10^{-1}$  & 5.69 (5.5 \%) & 5.78 (5.0 \%) & 5.70 (5.9 \%) & $\cdots$ & 2\% &  3\%\\
			$\delta/\pi$ & 1.52 (9.0 \%) & 1.58 (9.0 \%) &  1.54 (9.0 \%) & $\cdots$  &  13 \% &  {$\cdots$}\\
			[1pt]
			\hline\hline
		\end{tabular} \vspace{0.3cm}
		\caption{ The global fits and the prospects of neutrino oscillations parameters for IO. The first three columns summarize the results of three global-fit groups in the form of ``best fit ($1\sigma$ precision  in \%)''~\cite{Capozzi:2021fjo,deSalas:2020pgw,Esteban:2020cvm}. 
			The last three columns give the prospective precision of future neutrino oscillation experiments JUNO~\cite{JUNO:2022mxj}, DUNE~\cite{DUNE:2020jqi}, and HK~\cite{Hyper-Kamiokande:2018ofw}, respectively.
		}
		\label{Tab:OscParamAccuracy} 
	\end{table*}
	
	\begin{table}[b!]
		\centering
		\begin{tabular}{ccc|ccc}
			\toprule
			NME   & Ref.  & \Xe\ &  NME & Ref. &\Xe\ \\ 
			\hline\hline
			\multirow{2}{*}{IBM}    &  \cite{Barea:2015kwa} & 3.25   &   \multirow{3}{*}{EDF}  & \cite{Rodriguez:2010mn}   & 4.20    \\
			& \cite{Deppisch:2020ztt}         & 3.40    & & 
			\cite{Song:2017ktj}                & 4.24 \\ \cline{1-3}
			\multirow{5}{*}{QRPA}	& \cite{Mustonen:2013zu}                 & 1.55        & & \cite{ LopezVaquero:2013yji}                  & 4.77    \\ \cline{4-6}
			& \cite{Simkovic:2018hiq}               &2.72      &  	\multirow{3}{*}{NSM}  & \cite{Menendez:2017fdf} & (2.28, 2.45)   \\
			& \cite{Hyvarinen:2015bda}             & 2.91  &    	& \cite{Horoi:2015tkc}                  & (1.63, 1.76)          \\
			& \cite{Terasaki:2020ndc}             & 3.38     &   	& \cite{Coraggio:2020hwx}                  & 2.39       \\ \cline{4-6}
			& \cite{Fang:2018tui}                 & (1.11, 1.18)  & All & $\cdots$ & (1.11, 4.77) \\
			\hline \hline
		\end{tabular} \vspace{0.3cm}
		\caption{ \footnotesize The nuclear matrix elements $|M_{0\nu}|$ for \Xe\  evaluated with the IBM, QRPA, EDF and NSM methods, adapted from Ref.~\cite{Agostini:2022zub}.}
		\label{Tab:NME}
		
	\end{table}
	
	A simplified statistical analysis just counts the number of $\beta\beta$ events within the region of interest (ROI) without fitting the whole electron spectrum. This should be sufficient for our discussion at the sensitivity level.
	For a nominal $0\nu\beta\beta$ decay experiment with a total observed event number $N^{\rm exp}_{\rm tot}$, the likelihood of a hypothesis $H$ can be evaluated by assuming the Poisson fluctuation,
	\begin{eqnarray}\label{eq:PoissonD}
	{\mathcal{L}^{}_{0\nu} (N^{\rm th}_{0\nu}) }=  \frac{(N^{\rm th}_{0\nu}+B)^{N^{\rm exp}_{\rm tot}}}{N^{\rm exp}_{\rm tot}!} \cdot e^{-(N^{\rm th}_{0\nu}+B)} \;.
	\end{eqnarray}
	Here, $N^{\rm th}_{0\nu}$ is the theoretical expectation for the $0\nu\beta\beta$ decay event number of the hypothesis $H$, and $B$ is the background expectation.
	It is convenient to work on the basis of log-likelihood $-2 \ln{\mathcal{L}^{}_{0\nu}}$ which can be identified as the chi-square $\chi^2$ if the data sample is large and approximately Gaussian distributed.
	On top of the log-likelihood from $0\nu\beta\beta$ decays, we further add the information about neutrino masses and mixing angles from neutrino oscillation experiments, beta decays and cosmological observations, namely
	\begin{eqnarray}\label{eq:LogLH}
	-2 \ln{\mathcal{L}} & = & -2\ln{\mathcal{L}^{}_{0\nu}} -2 \ln{\mathcal{L}^{}_{\rm osc}} \\
	& & \hspace{1.cm} -2 \ln{\mathcal{L}^{}_{ \beta}} -2 \ln{\mathcal{L}^{}_{\rm cosmo}}  \;. \notag
	\end{eqnarray}
	The cosmological likelihood $\mathcal{L}^{}_{\rm cosmo}$ is obtained by analyzing the Markov chain file from the Planck Legacy Archive corresponding to the dataset {\it Planck} TT, TE, EE + lowE + lensing + BAO, which has set an upper bound on the sum of neutrino masses $\Sigma m^{}_{i} <0.12~{\rm eV}$~\cite{Aghanim:2018eyx}.
	The beta-decay likelihood is simply taken from Ref.~\cite{KATRIN:2021uub}. Note that the neutrino mass information for our choice is dominated by the cosmological observations.
	The likelihood of oscillation parameters is constructed by noting the equivalence $\chi^2_{\rm osc} = -2 \ln{\mathcal{L}^{}_{\rm osc}}$ assuming Gaussian distributions. Our neutrino oscillation chi-square $\chi^2_{\rm osc}$ is taken as 
	\begin{eqnarray}\label{eq:chi2}
	\chi^2_{\rm osc} \equiv \sum^{}_{i} \frac{(\Theta^{i}_{\rm osc} - \Theta^{{\rm bf},i}_{\rm osc})^2}{\sigma^2_{i}} \;,
	\end{eqnarray}
	where $\Theta^{i}_{\rm osc}$ include $\{\sin^2{\theta^{}_{13}}, \sin^2\theta^{}_{12}, \Delta m^{2}_{\rm sol}, \Delta m^{2}_{\rm atm}\}$, 
	$\Theta^{{\rm bf},i}_{\rm osc}$ are the best-fit values, and $\sigma^{}_{i}$ are the $1\sigma$ symmetrized errors.
	Our best-fit values and the error of $\sin^2{\theta^{}_{13}}$ are taken from NuFIT 5.1~\cite{Esteban:2020cvm} for IO, whereas the errors of $\{ \sin^2\theta^{}_{12}, \Delta m^{2}_{\rm sol}, \Delta m^{2}_{\rm atm}\}$ are taken to be the projections after 6-year running of JUNO~\cite{JUNO:2022mxj}.
	Note that by using Eq.~\eqref{eq:chi2} we have assumed the oscillation parameters to be independent in their experimental determination. In principle, their possible correlations (though mild in most cases) should be taken into account. Our results in this work would hence represent a conservative assessment, and the inclusion of correlations will strengthen more or less our major conclusions.

	For completion, we summarize in Table.~\ref{Tab:OscParamAccuracy} the current global fits and the prospects of neutrino oscillation parameters, in the assumption of IO. 
	The first three columns report the  best-fit values  along with their uncertainties (in parenthesis) from three major global-fit groups.
	The uncertainty of each parameter is represented by the $1\sigma$ fractional accuracy, defined as $1/6$ of the $3\sigma$ range divided by the best-fit value.
	The next three columns show the prospective precision of the upcoming neutrino oscillations experiments, i.e., JUNO \cite{JUNO:2022mxj}, DUNE \cite{DUNE:2020jqi}, and HK \cite{Hyper-Kamiokande:2018ofw}, respectively.

	Two different models can be compared with the help of likelihood ratio tests, as can be found in the textbook. The essence is to compute $\mathcal{L}^{\rm max}({H^{}_0})/\mathcal{L}^{\rm max}({H^{}_1})$ for two models $H^{}_{0}$ and $H^{}_{1}$, where the likelihood functions are maximized within each model by adjusting the model parameters. 
	A decent choice of the test statistic is simply given by
	\begin{eqnarray}\label{eq:DiffLogLH}
	-2\Delta\ln  \mathcal{L} \equiv -2\ln \frac{\mathcal{L}(H^{}_{0})}{\mathcal{L}(H^{}_{1})} \approx \chi^2(H^{}_{0})-\chi^2(H^{}_{1}) \;,
	\end{eqnarray}
	where the approximation of log-likelihood to a chi-square statistic holds when the sample size is large.
	This is not true in general for $0\nu\beta\beta$ decay searches, especially for those proposals with very low background indices, such as LEGEND, for which an $\mathcal{O}(1)$ signal event number can suggest a discovery.
	In such a case,
	the Monte-Carlo simulation with pseudo-experiments should be invoked to obtain the actual distribution of the test statistic $-2\Delta\ln  \mathcal{L}$ for the given model. The $p$-value to favor a model against another one can be obtained by counting the probability that the test statistic exceeds the experimentally observed $-2\Delta\ln  \mathcal{L}^{\rm obs}$.

	
	
	To be definitive, we fix the $0\nu\beta\beta$-decaying isotope as $^{136}{\rm Xe}$.
	The theoretical expectation of events $N^{\rm th}_{0\nu}$ severely replies on the input of NME of the decaying isotope.
	While the \textit{ab initial} approach is on its way,
	the existing phenomenological models available give rather diverse NME predictions, including, e.g.,
	the nuclear shell model (NSM), the quasi-particle random phase approximation (QRPA), the energy-density functional theory (EDF), and the interacting boson model (IBM).
	Different evaluations~\cite{Menendez:2017fdf,Horoi:2015tkc,Coraggio:2020hwx,Mustonen:2013zu,Hyvarinen:2015bda,Simkovic:2018hiq,Fang:2018tui,Rodriguez:2010mn,LopezVaquero:2013yji,Song:2017ktj,Barea:2015kwa,Deppisch:2020ztt,Terasaki:2020ndc} are summarized in Table.~\ref{Tab:NME} following Ref.~\cite{Agostini:2022zub}, which can differ by a factor more than four.
	For our numerical analysis, we treat the $^{136}{\rm Xe}$ NME in two ways:
	\begin{itemize}
		\item The NME uniformly (no preference in likelihood) distributes in the range $|M_{0\nu}| \in (1.11, 4.77)$.
		\item The NME takes a definitive value, e.g., $|M_{0\nu}| = 3$ with a negligible uncertainty, which can represent the ultimate goal of future theoretical evaluations.
	\end{itemize}
	As we will see quantitatively later, a precise knowledge of NME plays a vital role when we extract the neutrino information from the $0\nu\beta\beta$ decay data.
	%
	\begin{figure*}[t!]
		\begin{center}
			\subfigure{
				\includegraphics[width=0.4\textwidth]{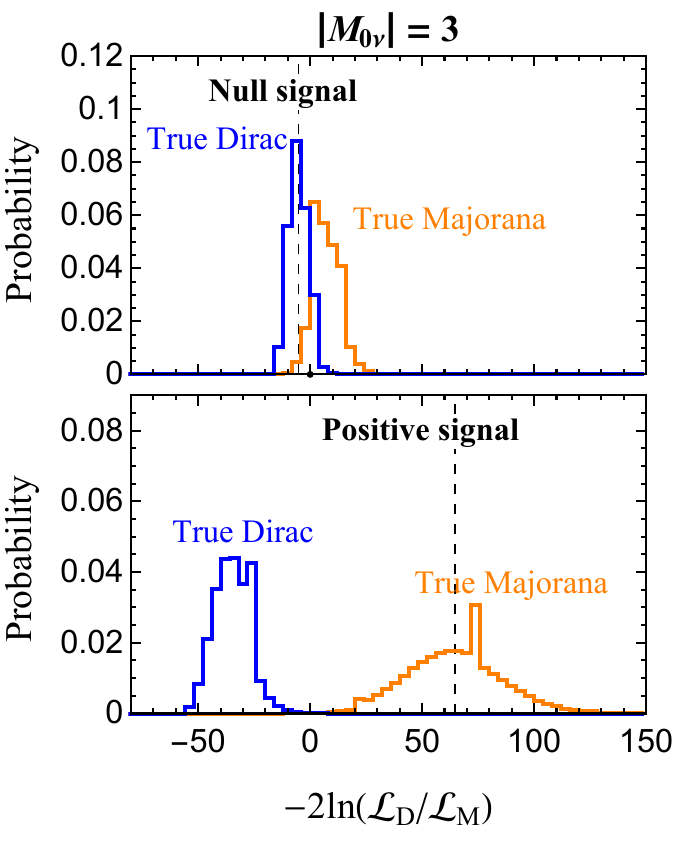} }
			\subfigure{
				\includegraphics[width=0.4\textwidth]{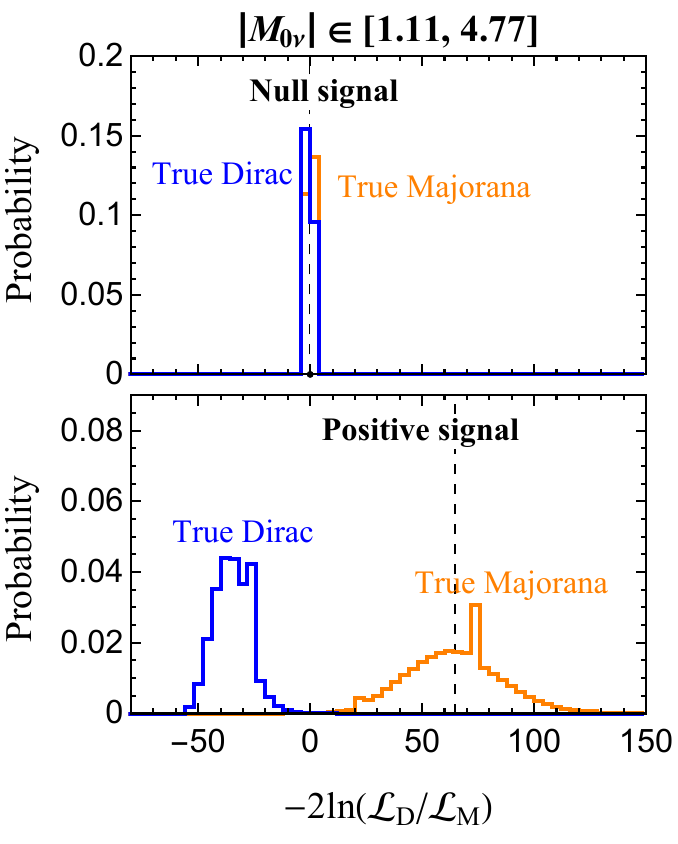} }
		\end{center}
		\vspace{-0.3cm}
		\caption{Probability distributions of the test statistic $-2\ln (\mathcal{L}^{}_{\rm D} / \mathcal{L}^{}_{\rm M})$  obtained from the Monte-Carlo simulations by assuming neutrinos are Dirac (blue histograms) or Majorana (orange histograms) fermions. 
			The experimental configuration is fixed as { $\xi = 5~{\rm ton \cdot yr}$ and $b = 1~{\rm  ton^{-1} \cdot yr^{-1}}$}.
			In the left panel, the NME is assumed to be completely known and fixed as $\left|{ M}^{}_{0\nu}\right| = 3$, while in the right panel, the NME takes a flat likelihood in the range of  $\left|{ M}^{}_{0\nu}\right| \in (1.11,4.77)$ during the fit.
			In each panel, the upper half one is obtained if a null signal has been observed (i.e., $N^{\rm exp}_{\rm tot} = B$) in a future $0\nu\beta\beta$ decay experiment, while the lower half stands for the scenario where a positive excess of events has been observed (i.e., $N^{\rm exp}_{\rm tot} - B >0$). The observed value of the test statistic is marked by the dashed  vertical line.
			In the case of a positive excess, we assume the observed signal event number to be $N^{\rm exp}_{\rm tot} - B \approx 27$ (can be read from Eq.~(\ref{eq:shl})), which is the expectation value by taking $m^{}_{3,{\rm true}}=0~{\rm eV}$, $\rho^{}_{\rm true} = 90^{\circ}$, and $\left|{ M}^{}_{0\nu}\right|^{}_{\rm true} = 3$.
			The model parameters of pseudo-experiments for all the hypotheses are located by maximizing the likelihood with respect to a null or positive signal.}
		\label{fig:MC-Simu}
	\end{figure*}

	The above considerations basically set up our statistical framework to compare models and to estimate model parameters.
	For the convenience of later discussions, we further establish the connection between a given experimental configuration and its sensitivity to $T^{0\nu}_{1/2}$.
	As is well known, the sensitivity of a generic $0\nu\beta\beta$ decay experiment to the half-life $T^{0\nu}_{1/2}$ can be estimated via the relation~\cite{Agostini:2017jim}
	\begin{eqnarray}\label{eq:shl}
	T^{0\nu}_{1/2} =\ln{2} \cdot \frac{N^{}_{\rm A} \cdot \xi \cdot \epsilon}{m^{}_{\rm iso} \cdot S^{}_{}(B)} \; ,
	\end{eqnarray}
	where $N^{}_{\rm A}$ is the Avogadro constant, $\xi$ represents the exposure, $\epsilon$ is the event detection efficiency, $m^{}_{\rm iso}$ is the molar mass of the specified isotope.
	Here, $S(B)$ represents the required expectation of signal event number within the ROI, for which a fraction $q$ of a set of identical experiments can report a discovery of $0\nu\beta\beta$ decays at a C.~L. larger than $p$.
	The dependence of $S(B)$ on $B = \xi \cdot b$ with $b$ being the background index is explicitly shown.

	\section{Inference from a null or positive signal}\label{sec:Result}
	\noindent
	Without otherwise specified, we will adopt an idealized experimental setup for the numerical assessment, assuming a background index $b=1~{\rm  ton^{-1} \cdot yr^{-1}}$ and a full detection efficiency $\epsilon = 100\%$. 
	The chosen background index corresponds to one count of background events for one ${\rm ton \cdot yr}$ of exposure, which represents an ultra-low background level achievable in the future.
	The only remaining experimental parameter, which is allowed to freely vary, is  the target exposure $\xi$.
	

	\begin{figure}[t!]
		\begin{center}
			\includegraphics[width=0.45\textwidth]{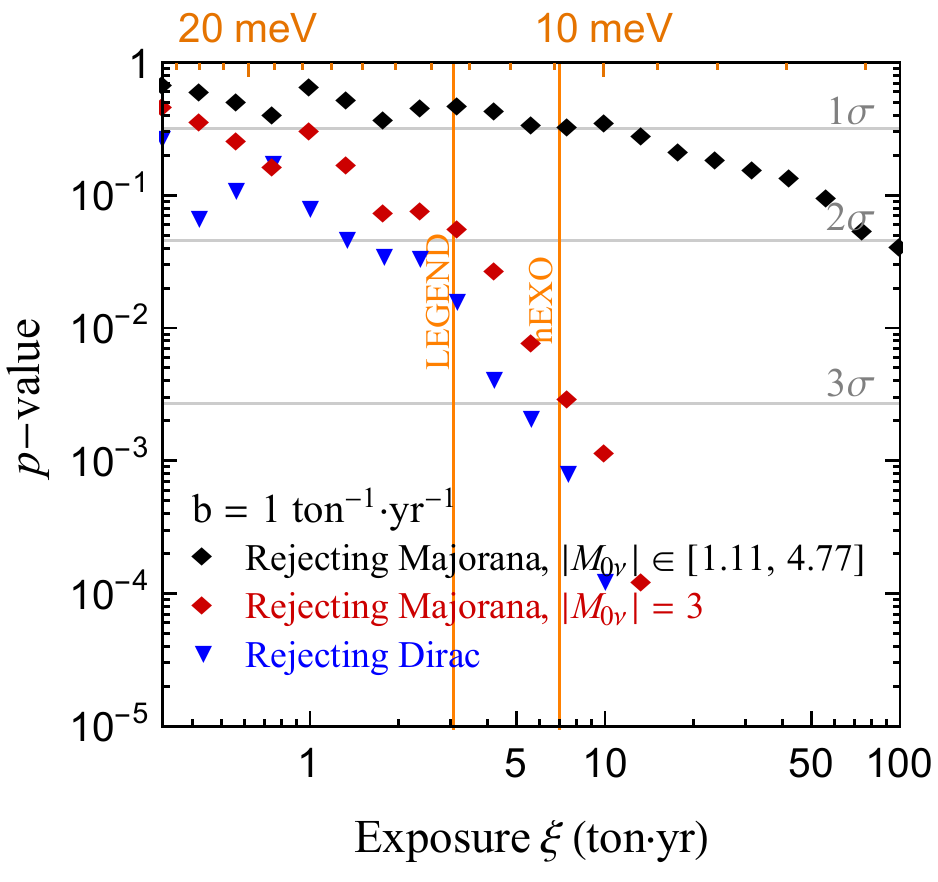} 
		\end{center}
		\vspace{-0.3cm}
		\caption{The significance level to reject the Majorana or Dirac neutrino hypothesis, if a null or positive signal has been observed, as a function of the $^{136}$Xe exposure. The positive signal is taken to be the expectation by choosing $m^{}_{3,{\rm true}}=0~{\rm eV}$, $\rho^{}_{\rm true} = 90^{\circ}$ and $\left|{ M}^{}_{0\nu}\right|^{}_{\rm true} = 3$ with other oscillation parameters in their NuFIT 5.1 best fits. 
			As usual, the background index is fixed as $b = 1~{\rm ton^{-1} \cdot yr^{-1}}$ with a full detection efficiency. The $90\%$ C. L. sensitivity to $|m^{}_{ee}|$ for the corresponding exposure is indicated on the top axis. The sensitivities of LEGEND~\cite{Abgrall:2017syy} and nEXO~\cite{Albert:2017hjq} to $|m^{}_{ee}|$ with averaged NMEs are shown as the vertical orange lines, which should be interpreted with the top axis.}
		\label{fig:p-value}
	\end{figure}
	
	\subsection{Dirac or Majorana nature}\label{sec:DorM}
	
	The first problem we want to address is to what extent the Majorana nature of neutrinos will be verified or falsified if the neutrino mass ordering turns out to be inverted.
	Following the statistical framework outlined in Sec.~\ref{sec:StatFrame}, our task is then to compare the Dirac neutrino hypothesis ${H}^{}_{\rm D}$ to the Majorana one ${H}^{}_{\rm M}$.
	There are two possible experimental outcomes for the next-generation $0\nu\beta\beta$ decay experiments:
	\begin{itemize}
		\item  {\it A null signal}. Namely, the observed event number $N^{\rm exp}_{\rm tot}$ coincides with the expectation of background $B$. During the fit, we first need to obtain the best-fit parameter values (e.g., $m^{}_{3}$, $\rho$ and $\theta^{}_{12}$) in ${H}^{}_{\rm D}$ and ${H}^{}_{\rm M}$, respectively, by maximizing the total likelihood in Eq.~(\ref{eq:LogLH}).
		The observed value of the test statistic $-2\Delta\ln  \mathcal{L}^{\rm obs} = -2\ln (\mathcal{L}^{}_{\rm D}/\mathcal{L}^{}_{\rm M}) $ is then calculated with the best-fit parameters within each model.
		To find out the $p$-value of the observed test statistic for a given hypothesis, we generate $-2\Delta\ln  \mathcal{L}$ as in  Eq.~(\ref{eq:DiffLogLH}) with pseudo-experiments assuming either ${H}^{}_{\rm D}$ or ${H}^{}_{\rm M}$ is true.
		An example of the probability distribution of $-2\Delta\ln  \mathcal{L}$ is illustrated in the upper two panels of Fig.~\ref{fig:MC-Simu}. 
		The observed $-2\Delta\ln  \mathcal{L}^{\rm obs}$ is marked by the vertical dashed lines.
		The experimental exposure is chosen to be $\xi = 5~{\rm ton \cdot yr}$.
		For the left panel, the NME is assumed to be completely known and taken as $|\mathcal{M}^{}_{0\nu}|=3$, and the $p$-value to accept the Majorana hypothesis is only $2\%$. 
		We notice that the distributions of Dirac and Majorana hypotheses are well separated (meaning that we can distinguish them well) if we have a plausible $|\mathcal{M}^{}_{0\nu}|$ calculation.
		Whereas, for the right panel, the NME is varying freely in the range $\left|{ M}^{}_{0\nu}\right| \in (1.11,4.77)$ during the fit, which severely dilutes the power of discrimination~\footnote{The fit of the null signal will always drive NME to the smallest possible value $|\mathcal{M}^{}_{0\nu}| = 1.11$.}, and a larger exposure is required to well separate them.

		\item {\it A positive signal}. One can do the exercise conversely, by assuming that a positive signal excess indicating Majorana neutrinos arises.
		To fix $N^{\rm exp}_{\rm tot}$ in this nominal analysis, 
		we first assume that neutrinos are Majorana fermions and choose a benchmark parameter choice for ${H}^{}_{\rm M}$:
		$m^{}_{3,{\rm true}}=0~{\rm eV}$, $\rho^{}_{\rm true} = 90^{\circ}$ and $\left|{ M}^{}_{0\nu}\right|^{}_{\rm true} = 3$ with other parameters in their NuFIT best fits.
		The resultant expectation of $N^{\rm th}_{0\nu}$ is then taken to be the observed signal event, i.e., $N^{\rm exp}_{\rm tot}- B =  27$ with $B=5$, a rather optimistic expectation with IO. 
		Similar to the null signal case, the Monte-Carlo results of the test statistic $-2\Delta\ln  \mathcal{L}$ are given in the lower two panels of Fig.~\ref{fig:MC-Simu}. 
		In this example, the distributions of Dirac and Majorana hypotheses are almost completely separated, and one is able to reject Dirac (or accept Majorana) with a very high significance.
		Interestingly, even in the case with an unknown NME, we can reject the Dirac hypothesis with a significance similar to the case with a fixed NME. 
		This is understandable as a positive signal will force the likelihood maximizer to find the true $\left|{ M}^{}_{0\nu}\right|$ for the Majorana hypothesis, in comparison to the case of null signal.
	\end{itemize}

	The above examples of discriminating Dirac and Majorana neutrinos have assumed a fixed experimental configuration.
	It is straightforward to explore the general sensitivity by varying the experimental exposure.
	We show in Fig.~\ref{fig:p-value} the $p$-value as a function of the experimental exposure $\xi$, in order to reject ${H}^{}_{\rm M}$ over ${H}^{}_{\rm D}$ with a null signal, and vise versa.
	The corresponding $1\sigma$, $2\sigma$ and $3\sigma$ significance levels of $p$-value are indicated by the  horizontal gray lines.
	%
	The red rhombus stands for the sensitivity to reject ${H}^{}_{\rm M}$ in the presence of the null signal by taking $\left|{ M}^{}_{0\nu}\right| =3$, while the black rhombus is for the varying NME $\left|{ M}^{}_{0\nu}\right| \in (1.11, 4.77)$.
	The blue triangle stands for the sensitivity to reject ${H}^{}_{\rm D}$ in the presence of the positive signal.
	The  event number of the positive signal is taken to be the expectation by choosing $m^{}_{3,{\rm true}}=0~{\rm eV}$, $\rho^{}_{\rm true} = 90^{\circ}$ and $\left|{ M}^{}_{0\nu}\right|^{}_{\rm true} = 3$, which gives $N^{\rm exp}_{\rm tot} - B \approx  27 \cdot \xi/(5~{\rm ton \cdot yr})$.
	It should be noted that this parameter choice is only used to generate the nominal events, since the actual $0\nu\beta\beta$ decay experiments of concern have not yet given any data.
	This is needed for the demonstrative purpose, but
	a different choice of the signal event number will more or less alter the results.
	After the nominal events are fixed, during the analysis we vary the relevant model parameters  freely and confine them with the total likelihood in Eq.~\eqref{eq:LogLH}. A different choice of $m^{}_{3,{\rm true}}$ in generating the nominal events does not make too much difference for the major results, because we can see from Fig.~\ref{fig:vissani} that the Planck experiment already pushed $m^{}_{3}$ to the regime where the dependence of $|m^{}_{ee}|$  on $m^{}_{3}$ becomes weak.

	\begin{figure*}[t!]
		\begin{center}
			\subfigure{
				\includegraphics[width=0.4\textwidth]{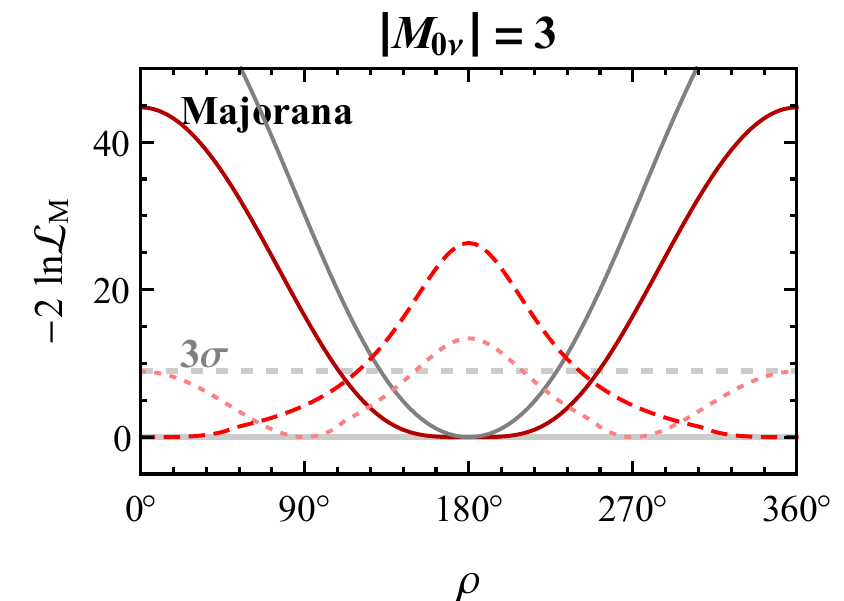} }
			\subfigure{
				\includegraphics[width=0.4\textwidth]{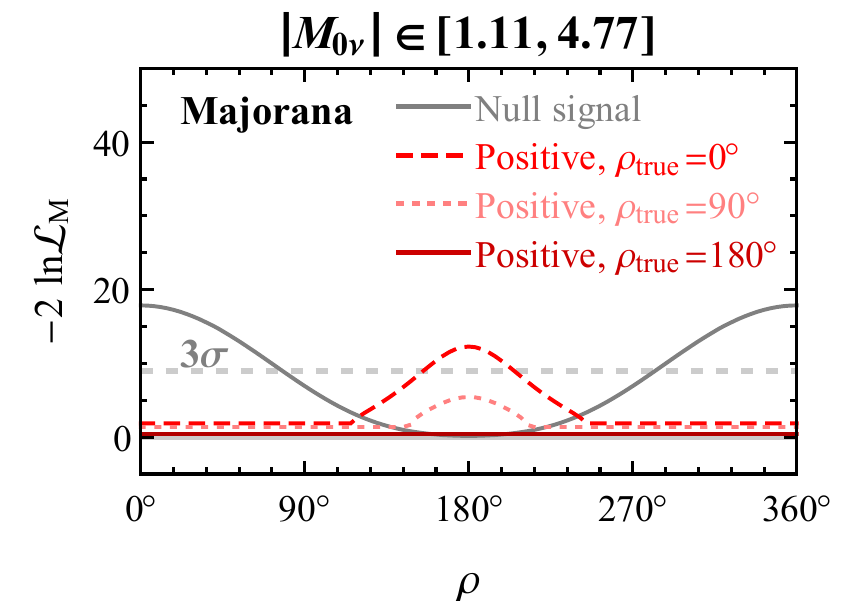} }
		\end{center}
		\vspace{-0.3cm}
		\caption{ The likelihood profile with respect to the Majorana CP phase $\rho$, by assuming $\left|{ M}^{}_{0\nu}\right| = 3$ (left panel) or $\left|{ M}^{}_{0\nu}\right| \in (1.11, 4.77)$ (right panel) during the fit.
			Here, the experimental configuration is fixed as { $\xi = 20~{\rm ton \cdot yr}$ and $b = 1~{\rm  ton^{-1} \cdot yr^{-1}}$}.
			The observed number of signal events $N^{\rm exp}_{\rm tot} - B$ is assumed to be vanishing (gray curves) or to be the expectation values of $m^{}_{3,{\rm true}} = 0~{\rm eV}$ and $\left|{ M}^{}_{0\nu}\right|^{}_{\rm true} = 3$ with $\rho^{}_{\rm true} = $ $0^{\circ}$ (dashed red  curves), $90^{\circ}$ (dotted pink curves) and $180^{\circ}$ (solid red curves), respectively.  }
		\label{fig:llh_rho}
	\end{figure*}

	\begin{figure*}[t!]
		\begin{center}
			\includegraphics[width=0.4\textwidth]{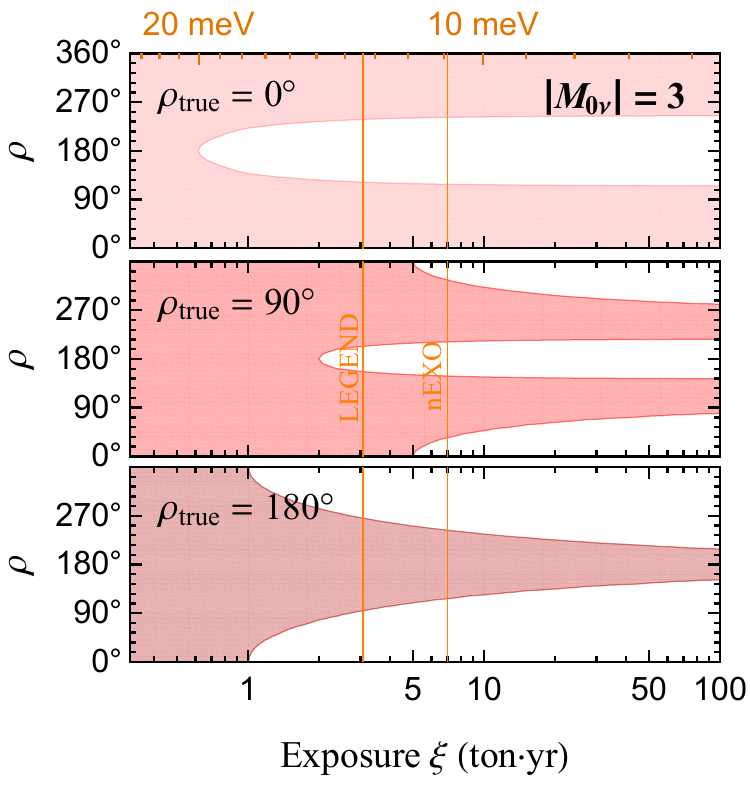} 
			\includegraphics[width=0.4\textwidth]{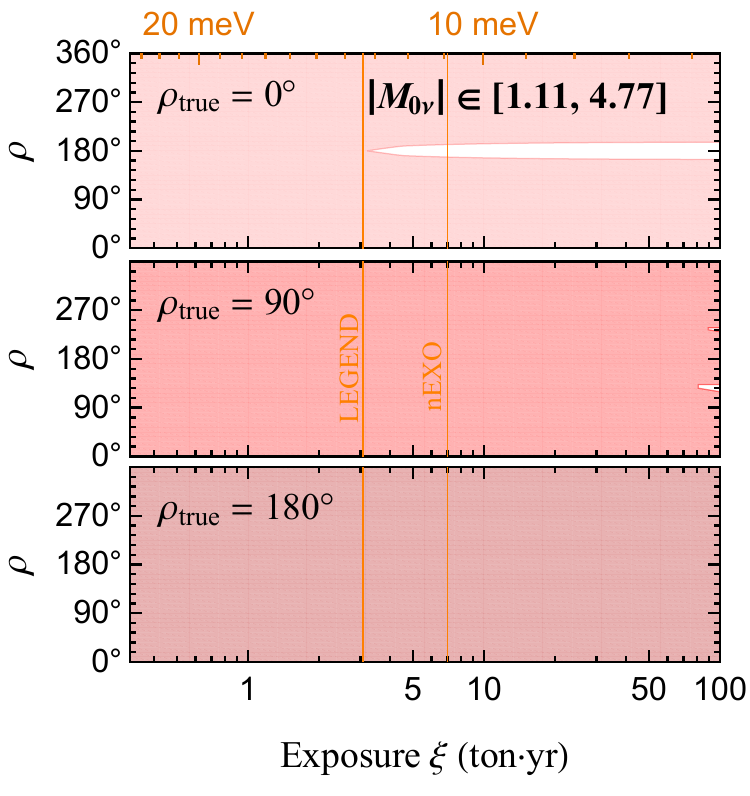} 
		\end{center}
		\vspace{-0.3cm}
		\caption{The $3\sigma$ sensitivities  to the Majorana CP phase $\rho$ with respect to the $^{136}$Xe exposure $\xi$, by assuming the NME to be completely known as $\left|{ M}^{}_{0\nu}\right| = 3$ (left panel), or spreading over a range $\left|{ M}^{}_{0\nu}\right| \in (1.11, 4.77)$ (right panel), during the fit. All the cases assume a positive signal of $0\nu\beta\beta$ decays. The signal event number takes the expectation value of the choice $m^{}_{3,{\rm true}} = 0~{\rm eV}$ and $\left|{ M}^{}_{0\nu}\right|^{}_{\rm true} = 3$ with $\rho^{}_{\rm true} = 0^{\circ}$ (top panel), $90^{\circ}$ (middle panel) or $180^{\circ}$ (bottom panel). }
		\label{fig:rho_xi}
	\end{figure*}

	In Fig.~\ref{fig:p-value}, the sensitivity to $|m^{}_{ee}|$ for the corresponding exposure with $b=1~{\rm  ton^{-1} \cdot yr^{-1}}$ is indicated on the top axis. 
	For a given exposure, this is obtained by converting $T^{0\nu}_{1/2}$ in Eq.~(\ref{eq:shl}) with $\left|{ M}^{}_{0\nu}\right| = 3$. We take the median sensitivity ($q = 50 \%$) to have a $90\%$ C. L. discovery potential.
	For comparison, the sensitivities of two projects, i.e., LEGEND ($8.5-19.4$ meV)~\cite{Abgrall:2017syy} and nEXO ($5.7-17.7$ meV)~\cite{Albert:2017hjq}, to $|m^{}_{ee}|$ with {\it averaged} NMEs are also indicated by the vertical lines.
	Note that  the LEGEND and nEXO lines should be interpreted with the help of the top axis. The exact capability of LEGEND and nEXO in distinguishing Dirac and Majorana hypotheses should have been performed by using their detailed experimental configurations. However, as we mentioned, to simplify the analysis we have adopted the framework in Sec.~\ref{sec:StatFrame} with the background index $b=1~{\rm  ton^{-1} \cdot yr^{-1}}$ and the full detection efficiency $\epsilon = 100\%$. In this case, the vertical lines should stand for the experiments with our idealized setup, which have equivalent $|m^{}_{ee}|$ sensitivities as the true LEGEND and nEXO experiments.

	Some further remarks are as follows.
	\begin{itemize}
		\item If the uncertainties in NMEs are negligible in the future and take the average of current theoretical evaluations ($\overline{\left|{ M}^{}_{0\nu}\right|} \approx 3$), the future projects such as LEGEND and nEXO can distinguish the Dirac and Majorana hypotheses with more than $2\sigma$ C. L. This can be seen by reading off the $p$-value of the red rhombus and blue triangle intersecting with the LEGEND and nEXO lines. The $2\sigma$ C. L. corresponds to a rejection $p$-value around $5\%$. 
		\item Pessimistically, one may presume that no further progress is achieved in the NME evaluation in the future. In this case, even with the sensitivity of nEXO one cannot tell whether neutrinos are Dirac or Majorana with a considerable significance, i.e., at most $1\sigma$, if the Dirac hypothesis is true. 
		Whereas, if the Majorana hypothesis is true, the discrimination significance will rely on how much the actual NME is.
		We do not necessarily need a precise knowledge of NME to reject the Dirac neutrino hypothesis if the actual NME is large, in which case the actual event excess will be large.
	\end{itemize}

	\subsection{Majorana CP phases}\label{sec:MajCP}
	If a positive signal excess is observed in the future, the most important message will be that neutrinos are Majorana fermions. Other than that, one can also attempt to extract additional unique information.  
	Because neutrino oscillation parameters have been well measured and will be further improved by JUNO in the very near future, the observable quantity $T^{0\nu}_{1/2}$ is mainly a function of the lightest neutrino mass $m^{}_{3}$, the CP phases and the NME $\left|{ M}^{}_{0\nu}\right|$.
	Furthermore, the sum of neutrino masses is severely constrained by the cosmological observations if no new physics is present, which limits the magnitude of $m^{}_{3}$ such that the dependence of $T^{0\nu}_{1/2}$ on $m^{}_{3}$ variation becomes weak.
	Hence, from the measurement of $T^{0\nu}_{1/2}$ one can simultaneously constrain the Majorana phase $\rho$ in the neutrino sector and $\left|{ M}^{}_{0\nu}\right|$ in the nuclear sector.
	%
	
	\begin{figure}[t!]
		\begin{center}
			\includegraphics[width=0.4\textwidth]{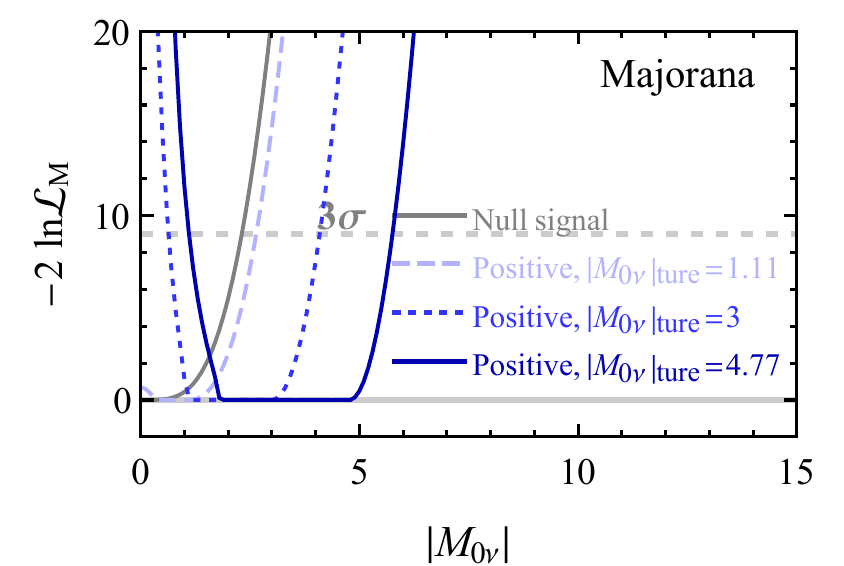} 
		\end{center}
		\vspace{-0.3cm}
		\caption{ The likelihood profile for the NME $\left|{ M}^{}_{0\nu}\right|$, in the presence of a null signal (gray curve) and a positive signal assuming $m^{}_{3,{\rm true}}=0~{\rm eV}$ and $\rho^{}_{\rm true} = 90^{\circ}$ with $\left|{ M}^{}_{0\nu}\right|^{}_{\rm true} = 1.11$ (dashed blue curve), $3$ (dotted blue curve) or $4.77$ (solid blue curve).	The experimental configuration is taken as $\xi = 20~{\rm ton \cdot yr}$ and $b = 1~{\rm  ton^{-1} \cdot yr^{-1}}$.}
		\label{fig:llh_M}
	\end{figure}
	
	\begin{figure}[t!]
		\begin{center}
			\includegraphics[width=0.4\textwidth]{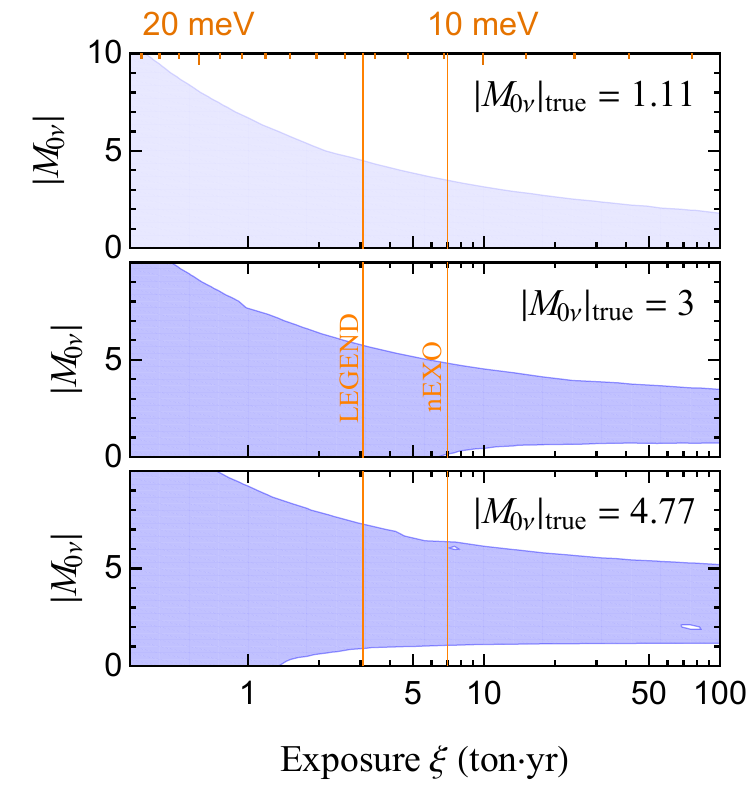} 
		\end{center}
		\vspace{-0.3cm}
		\caption{ The  $3\sigma$ sensitivities to the NME in terms of the $^{136}$Xe exposure $\xi$. 
			We assume that a positive signal has been observed in the experiment. The signal event number takes the expectation value of the choice $m^{}_{3,{\rm true}}=0~{\rm eV}$ and $\rho^{}_{\rm true} = 90^{\circ}$ with $\left|{ M}^{}_{0\nu}\right|^{}_{\rm true} = 1.11$ (top panel), $3$ (middle panel) or $4.77$ (bottom panel).}
		\label{fig:llh_Exposure}
	\end{figure}
	
	We start by exploring the sensitivity to $\rho$ by making assumptions about the theoretical input of $\left|{ M}^{}_{0\nu}\right|$.	
	In principle, we will need a positive $0\nu\beta\beta$ decay signal to set bounds on $\rho$, since a null signal can always be interpreted as a sign of Dirac neutrinos.
	Assuming a nominal event number $N^{\rm exp}_{\rm tot}$ has been observed, we can set constraints on $\rho$ as follows.
	First, we scan over all the model parameters and calculate the corresponding likelihood
	${\mathcal{L}} (\theta^{}_{13},\theta^{}_{12},\Delta m^{2}_{\rm sol},\Delta m^2_{\rm atm}, m^{}_{3}, \rho, \sigma, \left|{ M}^{}_{0\nu}\right| )$ with Eq.~(\ref{eq:LogLH}).
	Second, for every $\rho$ we marginalize over all the other parameters to obtain the maximum $\mathcal{L}$ (or the minimum $-2 \ln{\mathcal{L}}$).
	Finally, we obtain the likelihood profile $-2 \Delta\ln{\mathcal{L}}$ as a function of $\rho$ by subtracting its global minimum.
	The likelihood profiles of $\rho$ with various nominal inputs  are given in Fig.~\ref{fig:llh_rho} for the experimental exposure $\xi = 20~{\rm ton \cdot yr}$. 
	Four benchmark choices are illustrated: assuming $N^{\rm exp}_{\rm tot}-B$ to be vanishing (gray curves) or to be the expectation values of $m^{}_{3,{\rm true}} = 0~{\rm eV}$ and $\left|{ M}^{}_{0\nu}\right|^{}_{\rm true} = 3$ with $\rho^{}_{\rm true} = $ $0^{\circ}$ (dashed red  curves), $90^{\circ}$ (dotted pink curves) and $180^{\circ}$ (solid red curves), respectively.
	The $3\sigma$ C. L. range of $\rho$ can be obtained by setting $-2 \ln{\mathcal{L}^{}_{\rm M}} = 9$.
	In the absence of any signal events, one can still constrain $\rho$ because smaller $|m^{}_{ee}|$ will be preferred. However, as has been mentioned, we will be favoring Dirac over Majorana hypothesis in this case.

	As usual, we investigate two extreme scenarios of $^{136}{\rm Xe}$ NME, i.e., $\left|{ M}^{}_{0\nu}\right| = 3$ with a negligible uncertainty (left panel) and $\left|{ M}^{}_{0\nu}\right| \in (1.11, 4.77)$ (right panel).
	We find that the inclusion of the NME uncertainty will significantly dilute the sensitivity to $\rho$ due to the understandable degeneracy. This result further encourages the efforts from the nuclear community to push forward the plausible calculation of $\left|{ M}^{}_{0\nu}\right|$.

	The dependence of the sensitivity on the nominal exposure $\xi$ is shown in Fig.~\ref{fig:rho_xi}, where we give the $3\sigma$ allowed range of $\rho$ with a signal event number calculated from $m^{}_{3,{\rm true}} = 0~{\rm eV}$ and $\left|{ M}^{}_{0\nu}\right|^{}_{\rm true} = 3$ with $\rho^{}_{\rm true} = 0^{\circ}$ (top panel), $90^{\circ}$ (middle panel) or $180^{\circ}$ (bottom panel).
	In the left panel, the NME is assumed to be known as $\left|{ M}^{}_{0\nu}\right| = 3$, and next-generation $0\nu\beta\beta$ decay experiments will be able to rule out a significant amount of parameter space of the Majorana CP phase $\rho$.
	In comparison, for the right panel we observe that the lack of knowledge of  $\left|{ M}^{}_{0\nu}\right|$ significantly reduces the constraining power to $\rho$, even with an exposure of $100~{\rm ton \cdot yr}$ and an ultra-low background of $b=1~{\rm  ton^{-1} \cdot yr^{-1}}$.

	The measurement of Majorana CP phases is of great theoretical importance. For instance, 
	the $\mu{-}\tau$ reflection symmetry~\cite{Harrison:2002et} by far remains one of the best flavor symmetry which is consistent with the neutrino oscillation data. 
	This symmetry leads to very sharp predictions for the Majorana CP phases ($\rho, \sigma = $ $ 0^\circ $ or $ 180^\circ $)~\cite{Xing:2015fdg,Nath:2018zoi}.
	%
	%
	From the left panel of Fig.~\ref{fig:rho_xi}, we notice that if one of the solution is true (e.g., $ \rho_{\rm true} = 0^\circ$), the remaining one ($ \rho_{\rm true} = 180^\circ$)  can be ruled out at 3$ \sigma $   C. L.  with an exposure of only $1~{\rm ton \cdot yr}$. 
	However, this capability is washed out, if the NME uncertainty persists in the future. 
	In practice, the actual constraining power might be between those two extreme cases, with a reduced but not completely negligible NME uncertainty.

	\subsection{Nuclear matrix element}\label{sec:NME}
	The NME, as an imperative theoretical input, can be conversely ``measured'' by $0\nu\beta\beta$ decay experiments.
	If IO turns out to be true, the $0\nu\beta\beta$ decay will become an excellent ruler for the NME, though the resolution of this ruler is limited by the unknown CP phases in $|m^{}_{ee}|$.
	For NO, the uncertainty of $|m^{}_{ee}|$, mainly due to the ``well structure'', will be too large to extract any upper bounds on NME. Nevertheless, deriving a lower bound is always possible for NO with a positive observation of $0\nu\beta\beta$ decays, which is, however, not the main concern of the present work.

	Following a  procedure similar to fitting the CP phase, in Fig.~\ref{fig:llh_M} we generate the likelihood profile as a function of $\left|{ M}^{}_{0\nu}\right|$.
	For demonstration, several benchmark choices are shown, including a null signal (gray curve) and a positive signal assuming $m^{}_{3,{\rm true}}=0~{\rm eV}$ and $\rho^{}_{\rm true} = 90^{\circ}$ with $\left|{ M}^{}_{0\nu}\right|^{}_{\rm true} = 1.11$ (dashed blue curve), $3$ (dotted blue curve) or $4.77$ (solid blue curve).

	In Fig.~\ref{fig:llh_Exposure}, we present the $3\sigma$ sensitivities to the NME with respect to the nominal exposure $\xi$. The true value of NME is taken to be $\left|{ M}^{}_{0\nu}\right|^{}_{\rm true} = 1.11$ (top panel), $3$ (middle panel) or $4.77$ (lower panel).
	We have several remarks.
	\begin{itemize}
		\item In the given example, if the true value of $\left|{ M}^{}_{0\nu}\right|$ is $3$, a $3\sigma$ C. L. constraint $0.7 < \left|{ M}^{}_{0\nu}\right| < 3.5$ can be obtained with an exposure of $\xi = 100~{\rm ton\cdot yr}$, which is slightly narrower than the current prediction $\left|{ M}^{}_{0\nu}\right| \in (1.11, 4.77)$.
		\item  The resolution for NME is limited by the unknown Majorana CP phases. Hence, after the exposure increases to a certain level, the sensitivity will not be improved any further. Since there are not other foreseeable places to obtain the information of Majorana CP phases, this situation will persist regardless of the experimental achievements of $0\nu\beta\beta$ decays.
	\end{itemize}
	


	\section{Conclusion}\label{sec:conclusion}
	\noindent
	The next-generation $0\nu\beta\beta$ decay experiments are designed for a target sensitivity around $|m^{}_{ee}| \approx 10~{\rm meV}$, which sets the lower boundary of $|m^{}_{ee}|$ if the neutrino mass ordering is inverted.
	However, if NO is true and the lightest neutrino is unfortunately tiny, those next-generation projects will expect no $0\nu\beta\beta$ decay signals in the absence of new physics.
	While the mass ordering is still an open issue~\cite{Gariazzo:2022ahe} to be addressed by the upcoming neutrino oscillation experiments, we present here a frequentist analysis about what we can extract from $0\nu\beta\beta$ decays if IO eventually wins.
	From a positive or negative observation of $0\nu\beta\beta$ decays in the future, the physical information we can extract is at least three-fold.
	First and foremost, the nature of massive neutrinos can be well discriminated in the assumption of IO.
	Second, the Majorana CP phase $\rho$ can be probed with a positive $0\nu\beta\beta$ decay signal. Last, one can conversely constrain the NME in the presence of a signal.
	
	The role of NME in interpreting the unknowns in the neutrino sector is critical. 
	To efficiently probe the nature of neutrinos and the Majorana CP phases, a robust NME evaluation beyond the state of the art is necessary.
	Taking $^{136}{\rm Xe}$ for instance, if the actual NME is found to be $\left|{ M}^{}_{0\nu}\right| = 3$ with a negligible uncertainty, the experimental proposals like nEXO with a $10~{\rm meV}$ sensitivity can distinguish Dirac and Majorana neutrinos with $3\sigma$ C. L. or so, in the presence of a positive or null signal (as shown by Fig.~\ref{fig:p-value}).
	If $\left|{ M}^{}_{0\nu}\right|$ remains vague, for a null signal one cannot meaningfully reject the Majorana hypothesis with the next-generation design.
	Similar conclusions hold for the Majorana CP phases.
	The improved calculations of NMEs will make it possible to constrain one of the Majorana CP phases $\rho$ (see Fig.~\ref{fig:rho_xi}).
	In addition, in the worst case with a completely unknown NME, one can do the exercise conversely by constraining the NME with a positive $0\nu\beta\beta$ decay signal in the future (see Figs.~\ref{fig:llh_M} and \ref{fig:llh_Exposure}).
	
	%
	
	%
	
	

	\section*{Acknowledgements}
	\noindent GYH is supported by the Alexander von Humboldt Foundation. NN is supported by the Istituto Nazionale di Fisica Nucleare (INFN) through the “Theoretical Astroparticle Physics” (TAsP) project. 

	\bibliographystyle{utcaps_mod}
	\bibliography{KamLANDv1}

\end{document}